# The Symmetries and Origins of the Order Parameters for Pairing and Phase Coherence in Cuprates

A. Mourachkine

*Université Libre de Bruxelles, Service de Physique des Solides, CP233, Boulevard du Triomphe, B-1050 Brussels, Belgium*

The symmetries and origins of the order parameters (OPs) for pairing and long-range phase coherence in hole-doped cuprates are discussed. By analyzing tunneling, inelastic neutron scattering and torque measurements we conclude that the Cooper pairs in cuprates consist of spinons, and the pairing OP has an anisotropic s-wave symmetry, whereas the OP for long-range phase coherence has the magnetic origin due to spin fluctuations and, has the $d_{x^2-y^2}$ symmetry. All conclusions are based exclusively on experimental facts.

The superconductivity (SC) in metals requires the presence of the Cooper pairs and the phase coherence among them [1]. In the BCS theory for conventional SCs [1], the pairing mechanism and mechanism of establishment of the long-range phase coherence are identical: two electrons in each Cooper pair are attracted by phonons, and the phase coherence among the Cooper pairs is established also by phonons. Both the phenomena occur almost simultaneously at $T_c$. In SC copper-oxides, there is a consensus that, in the underdoped regime, the pairing occurs above $T_c$ when the long-range phase coherence is established, $T_{pair} > T_c$ [2-8]. Moreover, there are strong indications that the origins of the two mechanisms are different [8,3,4]. First of all, the magnitudes of the order parameters (OPs) for pairing and long-range phase coherence in hole-doped cuprates have different dependencies on hole concentration, $p$ in CuO$_2$ planes [2-4]. The magnitude of the OP for long-range phase coherence, $\Delta_c$, which is proportional to $T_c$, has the parabolic dependence on $p$ [3,4,9], while the magnitude of the OP for pairing, $\Delta_p$ increases linearly with the decrease of hole concentration [3,4,10]. Figure 1 shows the phase diagram for the two OPs in hole-doped cuprates [3]. In Fig. 1, the $\Delta_c$ scales with $T_c$ as $2\Delta_c/k_B T_c = 5.45$ [3]. It is worth noting that both the two gaps [11] are SC-like [8]. Secondly, an applied magnetic field affects the two OPs in cuprates differently: the long-range phase coherence disappears first, while in order to break the Cooper pairs the magnitude of magnetic field must be much higher [12]. However, in spite of the facts which show that the origins of the $\Delta_c$ and $\Delta_p$ are different, *there is no* consensus on the origins of these two OPs. The $\Delta_p$ gap occurs most likely due to pairing either spinons [2-4] or polarons [13,3]. Other scenarios for the origin of the $\Delta_p$ are practically implausible [2]. The $\Delta_c$ has most likely the magnetic origin [14,4]. Indeed, recently, there is the large number of experimental facts which show that spin fluctuations play a central role in hole-doped cuprates [15-18], and, as for bulk characteristics, the controlling factor in different cuprates seems only to be the hole concentration [15].

The origin and symmetry of a SC OP are tightly bound since the origin of a SC OP defines it's symmetry. For instance, in conventional SCs in which the SC is mediated by phonons [1], the SC OP has a s-wave symmetry. The magnetic origin of OP implies that the OP has the $d_{x^2-y^2}$ (hereafter, d-wave) symmetry [17]. In the present paper, we discuss simultaneously the symmetries *and* origins of the $\Delta_c$ and $\Delta_p$ in hole-doped cuprates. By analyzing inelastic neutron scattering (INS), torque and tunneling measurements we conclude that the $\Delta_p$ has an anisotropic s-wave symmetry and, is responsible for coupling of spinons, whereas the $\Delta_c$ originates from spin-spin exchange interactions and, has the d-wave symmetry.

There is a general consensus that the predominant OP in hole-doped cuprates has the d-wave symmetry [19]. This implies that one OP out of the two OPs has the d-wave symmetry. What is the symmetry of the second OP in hole-doped cuprates? In-plane torque anisotropy measurements on $Tl_2Ba_2CuO_{6+x}$ (Tl2201) show that the total OP consists of four-fold and two-fold components. The latter was interpreted by the presence of s-wave OP [20]. It is worth noting that a torque measurement which is sensitive only to the magnitude of a SC OP is a true *bulk* experiment [20]. Figure 2 shows the magnitudes of the two- and four-fold components obtained from the fit of torque data [20]. It is interesting that the magnitude of the s-wave OP is larger than the magnitude of the d-wave component. One should note that, in general, in two-gap scenario, the predominant character of one gap and it's magnitude do not relate to each other [21]. From Figs. 1 and 2, one can infer that the $\Delta_c$ has the d-wave symmetry, and the $\Delta_p$ has a s-wave symmetry. This is in a good agreement with the fact that the $\Delta_c$ has most likely the magnetic origin [4,14]. The presence of s-wave component is also found in $YBa_2Cu_3O_{6+x}$ (YBCO) [22,23] and $Bi_2Sr_2CaCu_2O_{8+x}$ (Bi2212) [24] by tunneling measurements and in $La_{2-x}Sr_xCuO_4$ (LSCO) by INS measurements [25]. Moreover, in LSCO, INS measurements revealed that the s-wave OP is responsible for pairing of spinons [25]. Thus, the set of torque, INS and tunneling measurements suggests that the $\Delta_p$ is responsible for pairing of spinons and, has a s-wave symmetry, while the $\Delta_c$ has the d-wave symmetry. Taking into account that the $\Delta_p$ has a s-wave symmetry and, *on the other hand*, (i) spin fluctuations are intrinsic in hole-doped cuprates studied so far [15-18]; (ii) the $\Delta_c$ has the d-wave symmetry, and (iii) the energy position of the so-called magnetic resonance peak coincides with the value of $2\Delta_c$ in hole-doped cuprates [14] and, is in quantitative agreement with the condensation energy [26], we conclude that the $\Delta_c$ has the magnetic origin due to spin fluctuations.

It is important to note that all conclusions made in the previous paragraph are based exclusively on experimental results. At the same time, all these conclusions are in a good agreement with a MCS (Magnetic Coupling between Stripes) model introduced recently [27,4]. The MCS model is based on a stripe model [2] which is in it's turn based on a spinon SC along charge stripes [28-30]. The main difference between the two models is that the coherent state of spinon SC is established differently in the two models, by spin fluctuations into local antiferromagnetic (AF) domains of $CuO_2$ planes in the MCS model, and by the Josephson coupling between stripes in the stripe model. Thus, in the MCS model, the SC has two different mechanisms: along charge stripes for pairing and perpendicular to stripes for establishing the coherent state. The pairing mechanism of spinons along charge stripes is described in Ref. 2. More detail information on the MCS model can be found elsewhere [12,27]. It is possible that the stripe model [2] is valid in the case of the SC in electron-doped $Nd_{2-x}Ce_xCuO_4$ (NCCO) cuprate in which the SC mediated by spin fluctuations is absent [4,27,12].

In fact, we are able also to find the approximate shape of the s-wave $\Delta_p$. The magnitude of total OP in Bi2212, which is equal, in general, to $\Delta = (\Delta_c^2 + \Delta_p^2)^{1/2}$ [31], has been measured by angle-resolved tunneling measurements [32]. Figure 3(a) shows tunneling data [32] obtained on Bi2212 single crystals with $T_c$ = 85 K ($p/p_m$ = 1.2). By knowing the magnitudes of the $\Delta_c$ and $\Delta_p$ at $p/p_m$ = 1.2 from Fig. 1, we find that, indeed, only the coherent gap $\Delta_c(1.2)$ = 20 meV may fit the data shown in Fig 3(a) *as the d-wave gap*. The $\Delta_p$ is too large for it. In Fig. 3(a), we present schematically the d-wave gap with the maximum magnitude of 20 meV. From $\Delta = (\Delta_c^2 + \Delta_p^2)^{1/2}$, we find the shape of the $\Delta_p$. Figure 3(b) shows *schematically* the shapes of the two OPs. Thus, the shape of the pairing s-wave OP in Bi2212 is anisotropic. Tunneling measurements on Pb-doped Bi2212 [33] support our findings that a tunneling gap with the maximum magnitude doesn't relate to $T_c$ (*i.e.* to $\Delta_c$). In the Pb-doped Bi2212 cuprate in which $CuO_2$ planes are, in the first approximation, unaffected by Pb-doping since Pb atoms reside between $CuO_2$ bilayers, the maximum of the gap magnitude which is located in Cu-Cu direction [see Fig. 3(a)] remains unchanged in comparison with pure Bi2212 case, while the minimum of the gap magnitude observed along Cu-O-Cu bond direction decreases proportionally to the decrease in $T_c$ [33]. Pb atoms doped in Bi2212 affect only the OP for long-range phase coherence keeping the pairing OP unchanged.

This implies that the pairing OP is present in Cu-Cu direction and the coherent OP is dominant in Cu-O-Cu bond direction. This is in a good agreement with Fig. 3(b).

One can argue that our conclusions contradict to angle-resolved photoemission (ARPES) measurements which show that the pairing OP has the d-wave symmetry [34]. However, the experimental data discussed above, which are in a good agreement among themselves, are much more reliable than ARPES data [35,36]. Even, if ARPES data are correct, they reflect the density of states (DOS) of the quasiparticle excitations in the thin layer of 3 Å on the surface [37] whereas, for example, torque measurements are true *bulk* experiments [20]. More information on this issue can be found elsewhere [36].

We turn now to the interpretation of INS spectra obtained on LSCO and YBCO. We found that by using our conclusions it is easy to interpret the INS spectra. First of all, it is important to note that an INS measurement is a signal averaging experiment. Secondly, one should notice that the resonance peak is the product of local AF domains of $Cu_2O$ planes and, appears at energy approximately equal to $2\Delta_c$ [14], whereas the spin gap occurs along charge stripes and, appears at $E \approx \Delta_p$. Thus, in an INS experiment, the two signals are overlapped. In the strongly underdoped regime where $\Delta_p > 2\Delta_c$, the resonance peak occurs inside the spin gap [16,38]. In slightly underdoped regime where $\Delta_p \approx 2\Delta_c$, the two signals are superimposed [39]. At optimal doping and in the overdoped regime where $\Delta_p < 2\Delta_c$, the resonance peak should appear above the spin gap. Figure 4 depicts odd and even INS spectra obtained on underdoped YBCO with $T_c$ = 52 K ($p/p_m$ = 0.54) at 5 K [38]. The broad peak at 63 meV in Fig. 4(a) corresponds to the anisotropic s-wave spinon gap $\Delta_p$ shown in Fig. 3(b), whereas the resonance peak at 25 meV corresponds to spin fluctuations in local AF domains of $Cu_2O$ planes. In YBCO, from the phase diagram shown in Fig. 1, the resonance peak at $p/p_m$ = 0.54 should be observed at 25 meV, and the spin gap $\Delta_p$ is equal approximately to 55 meV. Thus, there is a good agreement between the two sets of data. Moreover, in the optical mode shown in Fig. 4(b), the spin gap is observed at 53 meV [38]. In slightly underdoped LSCO ($x$ = 0.14), INS measurements show only one broad peak at around 20 meV [39]. In LSCO, from Fig. 1, the resonance peak at $p/p_m$ = 0.8 should appear at 17.5 meV, and the spin gap $\Delta_p$ is equal approximately to 19 meV. Thus, the two signals are superimposed, and this is in agreement with the INS data [39]. In slightly overdoped LSCO ($x$ = 0.163), INS measurements show the presence of almost isotropic s-wave spin gap with the magnitude of 6.7 meV and a broad peak at 11 meV. According to the phase diagram in Fig. 1, in slightly overdoped LSCO, the resonance peak should be observed at 18.2 meV, and the spin gap $\Delta_p$ is equal approximately to 13.4 meV. Unfortunately, the INS data in Ref. 25 are presented only up to 16 meV, and we can not discuss the presence of the resonance peak. However, it seems that the peak at 11 meV corresponds to the anisotropic s-wave $\Delta_p$ shown in Fig. 3(b) with the minimum and maximum magnitudes of 6.7 and 11 meV, respectively. In undoped $YBCO_{6.2}$, the resonance peak is obviously absent but the spin gap is detected in the optical mode at 67 meV [40], which is in a good agreement with Fig. 1. The shape of the resonance peak shown in Fig. 4 is unusually wide. In other INS measurements on underdoped YBCO [16], the shape of the resonance peak is very narrow, however, the broad peak corresponding to the spin gap is observed at energy which is somewhat larger than it should be according to Fig. 1. In fact, one should take into account that the phase diagram shown in Fig. 1 has to be used carefully, if the value of $p$ is calculated from the $T_c$ value. In addition to this, in YBCO, the situation is more complex than in other cuprates due to the presence of chains [4,27].

Finally, since theoretically, the d- and g-wave magnetically mediated SCs may co-exist [41] it is possible that, in hole-doped cuprates, there exists the g-wave OP [4,27,36] with the maximum magnitude of 30%-50% from the maximum magnitude of the d-wave component [41]. The g-wave component does not change the maximum magnitudes of the two OPs shown in Fig. 3(b), it just adds some "weight" between the maxima of the two OPs [36]. We understand that there is no consensus on the mechanism of the high-$T_c$ SC and origins of the $\Delta_c$ and $\Delta_p$. We therefore believe that our observations might trigger intense theoretical investigations and further experimental search to verify our conclusions.

In summary, we discussed the symmetries and origins of the OPs for pairing and long-range phase coherence in hole-doped cuprates. By analyzing tunneling, INS and torque measurements we concluded that the Cooper pairs in cuprates consist of spinons, and the pairing OP has an anisotropic s-wave symmetry, whereas the OP for long-range phase coherence has the magnetic origin due to spin fluctuations and, has the $d_{x^2-y^2}$ symmetry. We interpreted the INS spectra in YBCO and LSCO. We found that our conclusions which are based exclusively on experimental facts are in a good agreement with the MCS model of high-$T_c$ superconductivity.

I thank R. Deltour for discussion. This work is supported by PAI 4/10.

FIGURE CAPTIONS:

Fig. 1. Phase diagram in hole-doped cuprates at low temperature: $\Delta_c$ is the OP for long-range phase coherence, and $\Delta_p$ is the OP for pairing [3]. The $p_m$ is a hole concentration with the maximum $T_c$.

Fig. 2. Four- and two-fold components of total OP in Tl2201, obtained from the fit [20].

Fig. 3. (a) Tunneling gap vs. angle in Bi2212 at 5 K: dots (average measured points) and solid line (an assumption of a fourfold symmetry) [32]. The d-wave gap is shown schematically with the maximum magnitude of 20 meV (see text). (b) Shapes of two gaps at low temperature in overdoped Bi2212, shown schematically: the d-wave $\Delta_c$ and anisotropic s-wave $\Delta_p$.

FIG. 4. Averaged odd (a) and even (b) spin susceptibilities in YBCO at 5 K [38].

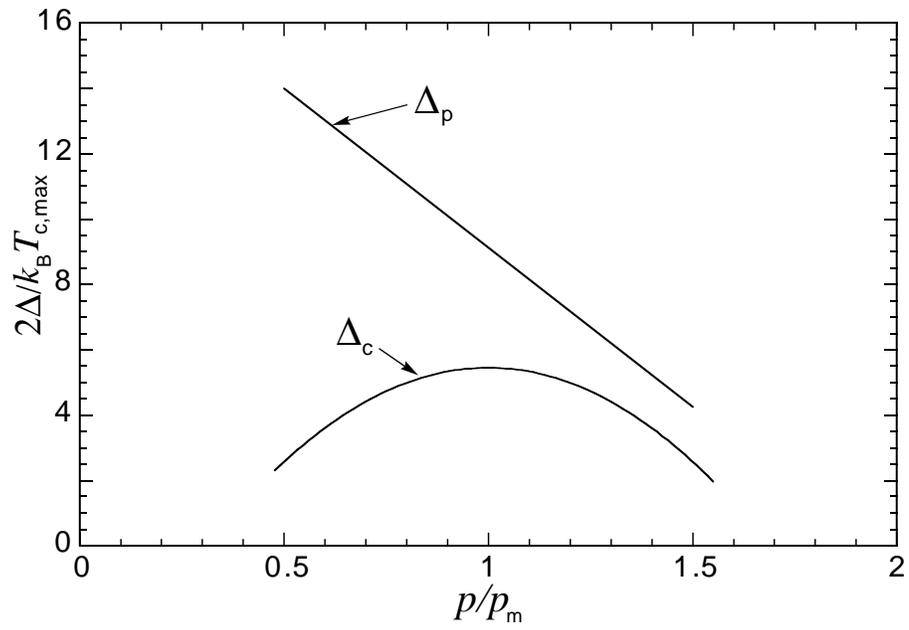

FIG. 1

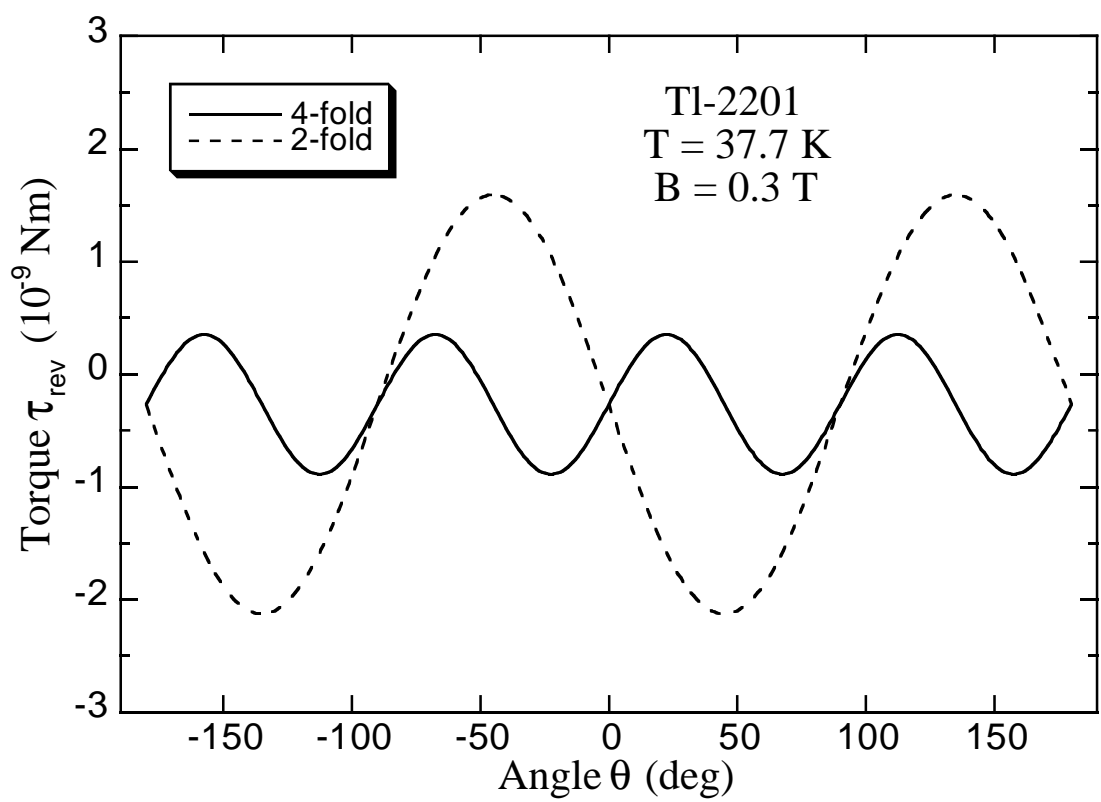

FIG. 2

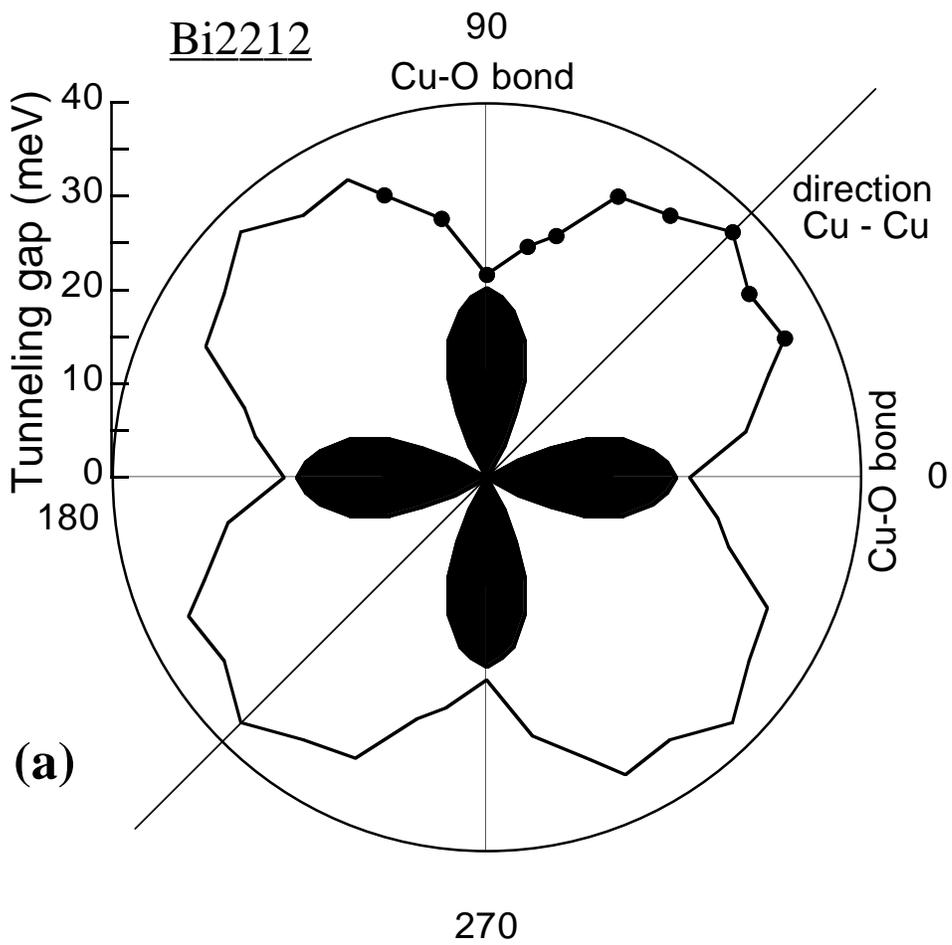

FIG. 3(a)

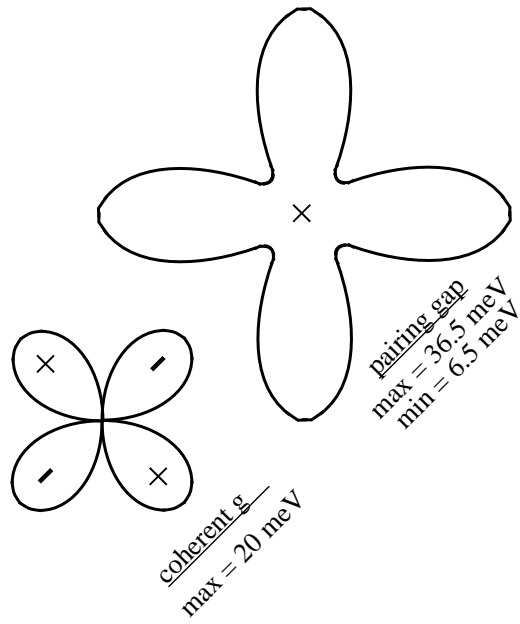

FIG. 3(b)

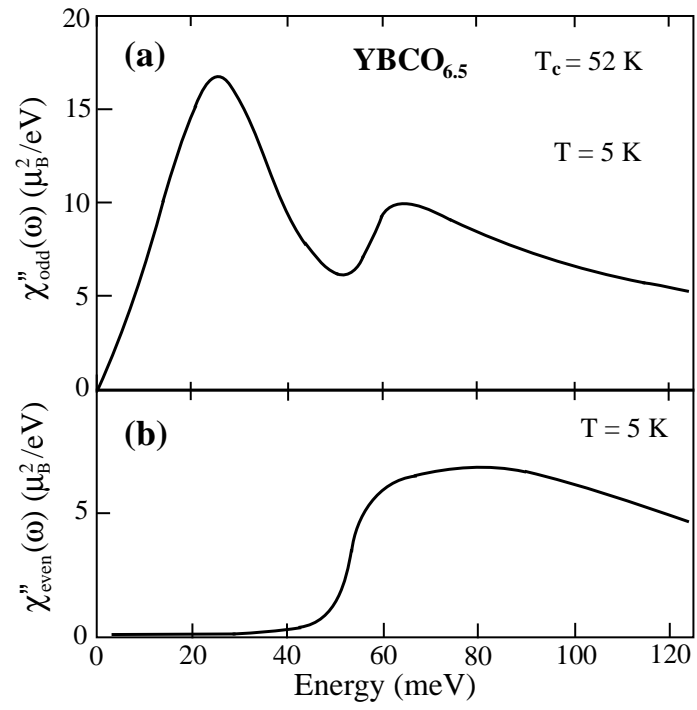

FIG. 4